      [1999/12/01 v1.4c Il Nuovo Cimento]
\documentclass{cimento}

\usepackage{graphicx}  % got figures? uncomment this
\title{Theoretical predictions of spectral evolution of short GRBs}
\author{F.~Fraschetti\from{ins:x},
R.~Ruffini\from{ins:x},
L.~Vitagliano\from{ins:x}
        \atque
S.S.~Xue\from{ins:x}}
\instlist{\inst{ins:x} ICRANet --- International Center for Relativistic Astrophysics and Dipartimento di Fisica, Universit\`a di Roma ``La Sapienza''}

\PACSes{\PACSit{95.30}{Astrophysical plasmas} \PACSit{98.70}{Gamma-Ray Bursts}}
\begin{document}

\maketitle

\begin{abstract}
We present the properties of spectrum of radiation emitted during gravitational collapse in which electromagnetic field strengths rise over the critical value for $e^+e^-$ pair creation. A drift from soft to a hard energy and a high energy cut off have been found; a comparison with a pure black body spectrum is outlined.\end{abstract}

The idea that the origin of GRBs is related to the energy extractable from a black hole \cite{CR71,RV02} by process of vacuum polarization \cite{S31} and the creation of $e^+e^-$ plasma was advanced in 1974 by \cite{DR75}. In 1998 \cite{PRX98} these concepts were further evolved by the identification of the region around an already formed black hole in which such $e^+e^-$ plasma can be created and the concept of ``dyadosphere'' was introduced.

The crucial issue
of the survival of the electric charge of the collapsing core in the presence of a
copious process of $e^+e^-$ pair creation was addressed in
\cite{RVX03a}. By using theoretical techniques borrowed from
plasma physics and statistical mechanics (see \cite{KME98} and references therein) based on a
generalized Vlasov equation, it was possible to show that while the core keeps
collapsing, the created $e^+e^-$ pairs are entangled in the
overcritical electric field. The electric field itself, due to the back
reaction of the created $e^+e^-$ pairs, undergoes damped
oscillations in sign finally settling down to the critical value
$\mathcal{E}_{\mathrm{c}}$. The pairs fully thermalize to an
$e^+e^-$-photon plasma on time-scales typically of the order of
$10^{2}$--$10^{4}\hbar/m_{e}c^{2}$. During this characteristic damping time,
which is much larger than the pair creation time-scale
$\hbar/m_{e}c^{2}$, the core moves
inwards, collapsing with a speed $0.2$--$0.8c$,
further amplifying the electric field strength at its surface and
enhancing the pair creation process. Meanwhile $e^+e^- \gamma$ plasma relativistically expands outwards until transparency when photons are emitted \cite{RSWX99}. We discretize the outgoing plasma in subshells and numerically integrate the hydrodynamics equations until the transparency condition is reached for every subshell.

%%% Figures:
\begin{figure}
\begin{center}
\includegraphics[width=8.5cm]{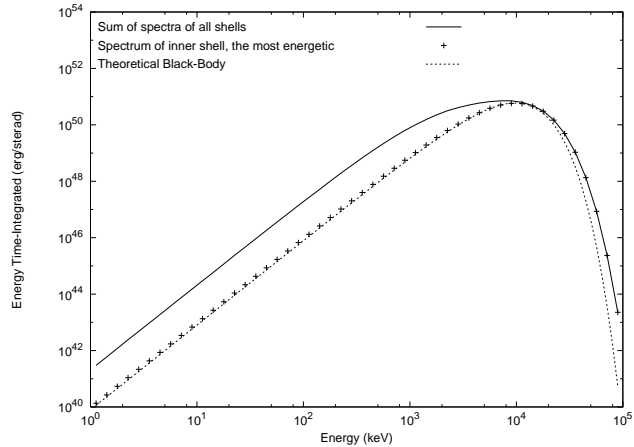}     % includes figure foo.eps
\caption{The theoretical prediction of total time integrated spectrum
for a short burst with $M=10M_{\odot}$ and $Q=0.1\sqrt{G}M$ is compared
with the spectrum of the inner shell, which is the most energetic. 
The contribution of all the other shells is to shift to
lower energies the energy peak and to broaden the curve. The spectrum of a pure black body
is also reported.}
\end{center}
\end{figure}

The results agree very closely with the observations of
short-bursts \cite{lett2,P...99,c06,v06}. A peculiar soft-to-hard evolution of spectral hardness has been found. Compared with the black body spectrum, three features of the time-integrated spectrum of our theoretical model are:
\begin{itemize}
\item in the low energy regime, the spectrum, being a superposition of black bodies at different temperatures, has the same slope as the black body one \cite{gcg03};
\item the spectrum is broader around energy peak, compared with black body spectrum; 
\item in the high energy regime, the spectrum is more sharply cut off due to separatrix \cite{RVX03a}, compared the exponential cut off of the black body spectrum. 
\end{itemize}
Evidence for the existence of an exponential cut off at high energies in the spectra of short bursts has been found \cite{g03}.

\end{document}